\documentclass{aa}  

\usepackage{graphicx}
\usepackage{txfonts}
\usepackage{aalongtable}
\begin{document} 
\setcounter{table}{1}

   \title{A GIANO-TNG high resolution IR spectrum of the airglow emission}


   \author{E. Oliva \inst{1}   
          \and L. Origlia \inst{2}  
	  \and R. Maiolino \inst{3} 
	  \and C. Baffa \inst{1}
          \and V. Biliotti \inst{1}
          \and P. Bruno \inst{4} 
          \and G. Falcini  \inst{1}
          \and V. Gavriousev \inst{1}
	  \and F. Ghinassi \inst{5} 
          \and E. Giani\inst{1}  
          \and M. Gonzalez\inst{5}
          \and F. Leone\inst{6}   
          \and M. Lodi\inst{5}
          \and F. Massi\inst{1}
          \and P. Montegriffo\inst{2}
	  \and I. Mochi\inst{7} 
          \and M. Pedani\inst{5}
          \and E. Rossetti\inst{8} 
          \and S. Scuderi\inst{5}
          \and M. Sozzi\inst{1}
          \and A. Tozzi\inst{1}
	  \and E. Valenti \inst{9} 
          }

   \institute{
              INAF - Osservatorio Astrofisico di Arcetri,
              Largo E. Fermi 5, I-50125, Firenze, Italy 
              \email{oliva@arcetri.inaf.it}
         \and
	     INAF - Osservatorio Astronomico di Bologna,
             via Ranzani 1, I-40127 Bologna, Italy
         \and
              Cavendish Laboratory, University of Cambridge,
              19 J.\ J.\  Thomson Avenue, Cambridge CB3 0HE, UK
	 \and
	     INAF - Osservatorio Astrofisico di Catania,
	      via S. Sofia 78, I-95123 Catania, Italy
	 \and
            INAF - Fundaci\'{o}n Galileo Galilei,
             Rambla Jos\'{e} Ana Fern\'{a}ndez P\'{e}rez, 7,
             38712 Bre\~{n}a Baja, TF, Spain
	 \and
            Dipartimento di Fisica e Astronomia, Sezione Astrofisica,
            Universit\`{a} di Catania,
	      via S. Sofia 78, I-95123 Catania, Italy
         \and
             Lawrence Berkeley National Laboratory,
             MS 2-400, One Cyclotron Road, Berkeley, CA 94720, USA
         \and
             Dipartimento di Astronomia, Universit\`{a} di Bologna,
             Via Ranzani 1, I-40127 Bologna, Italy
         \and
         European Southern Observatory,
         Karl-Schwarzschild-Str. 2, D-85748 Garching bei Muenchen, Germany
             }

   \date{Received .... ; accepted ...}

 
  \abstract
{}
{
A flux-calibrated high resolution spectrum of the airglow emission is
a practical $\lambda$-calibration reference for astronomical
spectral observations. 
It is also useful for
constraining the molecular parameters of the OH molecule and the
physical conditions in the upper mesosphere.
}
{
We use the data collected during the first technical commissioning
of the GIANO spectrograph at the Telescopio Nazionale Galileo (TNG).
The high resolution (R$\simeq$50,000) spectrum simultaneously covers the
0.95-2.4 $\mu$m wavelength range. Relative flux calibration is achieved by
the simultaneous observation of a spectrophotometric standard star.
}
{
We derive
a list of improved positions and intensities of OH infrared
lines. The list includes $\Lambda$-split doublets
many of which are spectrally resolved.
Compared to previous works, the new results correct errors in the 
wavelengths
of the Q-branch transitions.
The relative fluxes of OH lines from different vibrational bands
show remarkable deviations from theoretical predictions: 
the $\Delta v$=3,4 lines are a factor of 2 and 4 brighter than
expected.
We also find evidence of a significant fraction (1-4\%) of OH molecules
with ``non-thermal'' population of high-J levels.
Finally we list wavelengths and fluxes of 153 lines not attributable
to OH. Most of these can be associated to O$_2$, while 37 lines in the H
band are not identified.
The O$_2$ and unidentified lines in the H band account for $\simeq$5\%\
of the total airglow flux in this band.
}
{}


   \maketitle

%

   \begin{figure*}
   \includegraphics[width=\hsize]{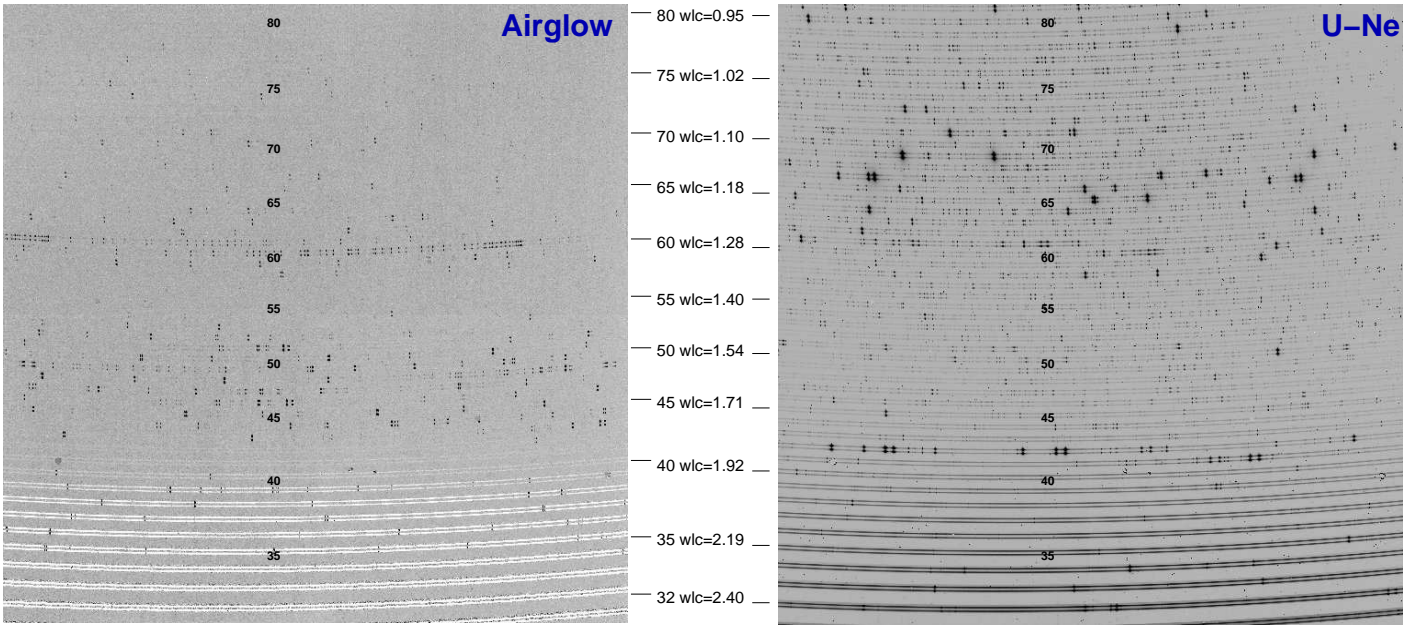}
      \caption{
           GIANO echelle spectra of the airglow emission (left panel)
           and of the U-Ne calibration lamp (right-hand panel). The
           positions of several orders and the corresponding central
           wavelengths (in $\mu$m) are marked.
              }
         \label{fig_2D_all}
   \end{figure*}

\section{Introduction}

The airglow emission is an annoying, unavoidable contamination of all
ground-based astronomical observations. 
It mostly consists of narrow lines 
of molecular bands which, on the other hand,
could be conveniently used as a reference spectrum
for wavelength calibration of spectroscopic data. 
The airglow lines at wavelengths 0.3--1.0 $\mu$m have been thoroughly 
compiled and modelled using high resolution (R$\simeq$10$^5$)
data from HIRES-Keck and UVES-VLT;
see e.g. Osterbrock et al. \cite{Osterbrock98}, Hanuschik \cite{Hanushik}, 
Cosby et al.  \cite{Cosby2006}.
The data in the near infrared are much sparser and based on lower 
resolution spectra, see e.g. Oliva \& Origlia \cite{Oliva92}, Maihara et al.
\cite{Maihara}. The most recent and complete line compilation is that
of Rousselot et al. \cite{Rousselot00} (hereafter R2000) 
which used ISAAC-VLT spectra at 
resolving power R$\simeq$8,000. Besides the relatively low resolution, which
blends many lines, these spectra have a very limited simultaneous wavelength
coverage $\Delta\lambda/\lambda$=1/16. Therefore, they cannot be used 
to measure intensity ratios of a sufficiently large sample of lines, 
because the airglow intensity changes between the different exposures needed
to cover the whole wavelength range. The ideal instrument for this type
of measurement is a cross-dispersed spectrograph, which can combine 
high spectral resolution and broad wavelength coverage. 
GIANO-TNG is the first instrument of this type available to the astronomical 
community. 

We present and discuss measurements taken with GIANO during its first
technical commissioning at Telescopio Nazionale Galileo (TNG). In section 
\ref{observations} we briefly describe the instrument, the measurements 
and the data reduction. In section \ref{results} we present and discuss
the results.

\section{Observations}
\label{observations}
GIANO is a cross-dispersed spectrograph which produces, in a single exposure,
a spectrum extending from 0.95~$\mu$m to 2.4~$\mu$m at a resolving power 
R$\simeq$50,000.  
The main disperser is a commercial R2 echelle grating
with 23.2 lines/mm which works in quasi-Littrow configuration on
a $d=100$ mm collimated beam. Cross dispersion is achieved via a network 
of fused silica and ZnSe prisms which work in double pass, i.e. they
cross-disperse the light both before and after it is dispersed by
the echelle gratings.
This setup produces a curvature of the images of the spectral orders.
More technical details on the instrument can be found in 
Oliva et al.\ ~\citep{oliva_spie2012_1,oliva_spie2012_2}        
and references therein.

The spectral layout on the detector is shown in Fig.~\ref{fig_2D_all}.
The echellogram on the detectors spans 49 orders, from \#32 to \#80.
The spectral coverage is complete up to 1.7 $\mu$m.
At longer wavelengths the orders become wider than the 
detector. The effective spectral coverage in the K-band is about 75\%.
The sky spectrum in the left panel of Fig.~\ref{fig_2D_all} 
was dark-subtracted
using an exposure taken with a blocking filter at room temperature, for this reason the
thermal continuum beyond 2 $\mu$m results in absorption.

For its first technical commissioning at the TNG, we used 
a bundle of 2 IR-transmitting ZBLAN fibres provisionally
connected to the TNG focus for visiting instruments.
These fibres are standard off-the-shelf products with a core of 85~$\mu$m,
 which corresponds to a sky-projected angle
of 1~arcsec.  The two fibres are aligned and mounted inside a custom connector.
The cores are at a distance of 0.25~mm, equivalent to a sky
projected angle of about 3~arcsec.
Due to the constraints set by the visitor focus, 
the fibre entrances were coupled to the TNG using a provisional, simplified
focal adapter which consisted of a commercial CaF$_2$ singlet lens positioned
26~mm before the fibres. 
The focal adapter was mechanically mounted at a fiducial position, no
further adjustment of the optical axis was possible. Unfortunately this
resulted in a very reduced efficiency of the system, which severely
limited the use of the instrument for observations of faint targets.
A pellicle beam-splitter, positioned just before the lens, was used to deviate
$\sim$8\% of the light to the guider CCD camera, working in the Z-band.
Light from calibration lamps could be fed into the fibres by inserting
a mirror in place of the beam-splitter.

The data were collected during part of the technical
nights from July 27$^{th}$ to July 30$^{th}$ {\bf 2012}. Sky-only spectra, such as
those shown in Fig.~\ref{fig_2D_all}, were collected by pointing at blank sky
positions. Sky+star spectra were collected by centring a hot star with
known flux (Hip89584) in one of the two fibres. These spectra were used to 
measure the relative fluxes of the airglow lines. 
The flux calibration did not include correction for telluric absorption
features, because the lines are not resolved. 
The geometry of the orders was determined using flat exposures with a 
tungsten calibration lamp. The 2D spectrum was thus rectified and the
spectra were extracted by summing 6 pixels around each fibre, in the 
direction perpendicular to dispersion.
Wavelength calibration was determined feeding the fibres with the light from
a U-Ne lamp. The wavelengths of the Uranium lines were taken from Redman
et al.~\cite{redman2011} while for Neon we used the table available
on the NIST website\footnote{physics.nist.gov/PhysRefData/ASD/lines\_form.html}.
The $\lambda$ vs. pixel relationship was obtained starting from a 
physical model 
of the instrument. This procedure is part of the pipeline that we are
developing for the instrument. 
The resulting wavelength accuracy was about $\lambda/300,000$
r.m.s., i.e. 0.05~\AA\  for lines in the H-band.

Relative flux calibration was performed approximating the photon flux of the
standard star (Hip89584) with the following interpolation formula
$$ \log(N_\lambda) = -3.46 -1.58 \log\lambda -1.93 (\log\lambda)^2 $$
where $\lambda$ is in $\mu$m and $N_\lambda$ is in photons/cm$^2$/s/$\mu$m.
The accuracy of the measured flux of bright lines in regions free of 
telluric absorptions is 10\%\ r.m.s.

\section{Results}
\label{results}
A total of about 750 airglow lines were detected in our spectra.
About 500 can be attributed to OH roto-vibrational transitions, 114 can be
associated to O$_2$, while the others are mostly not identified.

We first concentrate on the OH lines,
for which a rich theoretical background exists in the literature.

   \begin{figure*}
   \centering
   \includegraphics[angle=-90,width=0.85\hsize]{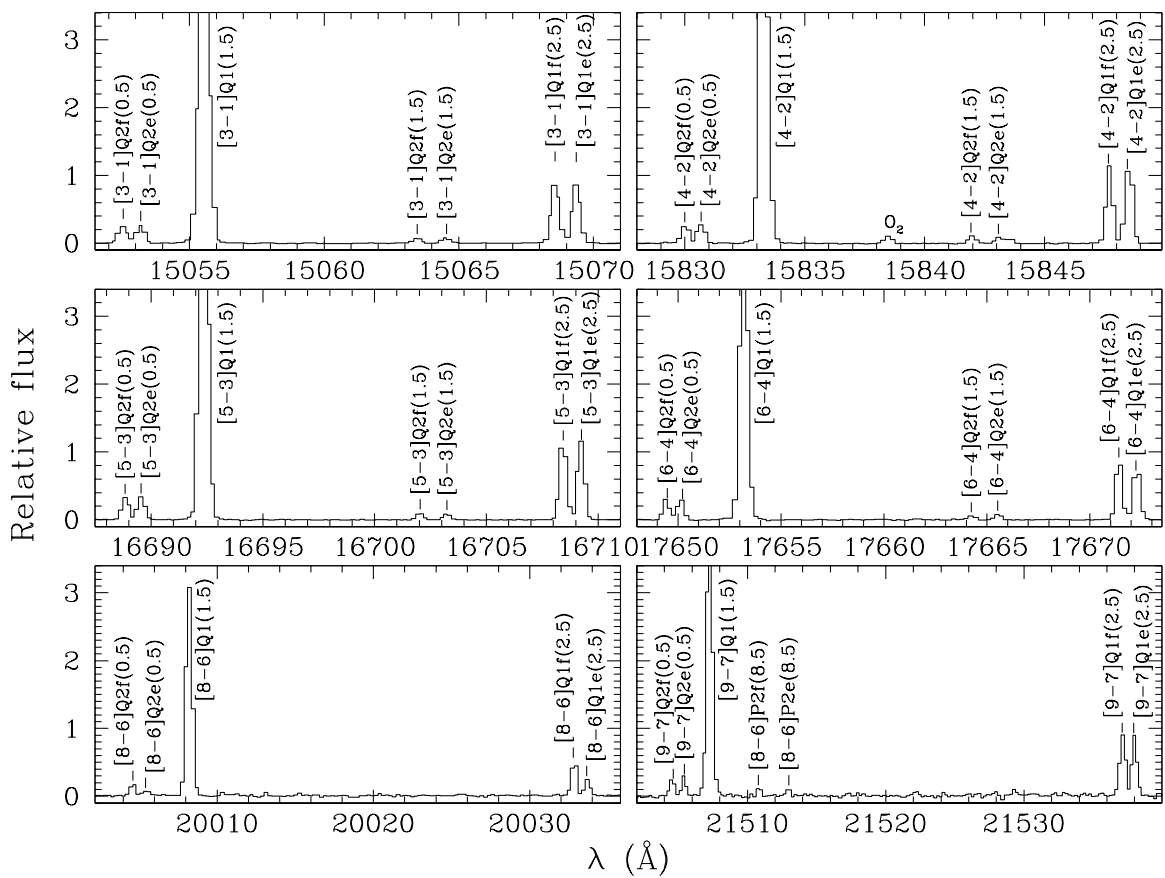}
      \caption{
    Parts of the extracted spectrum including the Q-branch lines.
    Most of the $\Lambda$-doublets are clearly resolved at the GIANO
    resolution.
              }
         \label{fig_Q_details}
   \end{figure*}

\subsection{ OH wavelengths and fluxes }
\label{OH_wavelengths_fluxes}

Table~\ref{table_oh_1} includes
the lines which were unambiguously identified as OH transitions.
For each $\Lambda$-doublet of OH lines we give the wavelengths (in vacuum)
and the total flux of the doublet. The relative intensities of the two
$e$,$f$ lines of each doublet, when resolved, were in all cases found equal
to unity, within the errors.

The listed wavelengths are derived from the OH energy level positions of
Abrams et al.~\cite{Abrams1994}. These are the most accurate molecular
data available in the literature and yield line wavelengths with an r.m.s.\ 
accuracy of 0.0035~cm$^{-1}$, equivalent to 0.01~\AA\  at 16,000~\AA.
Compared to the work of R2000, we find a discrepancy in the positions
of all the Q-lines. Specifically, we confirm that the $\Lambda$-doublets of
most of these lines
are clearly resolved at the GIANO resolution (see Fig.~\ref{fig_Q_details}).
In contrast to our results, 
the R2000 list predicts that these lines should be unresolved.
Similar discrepancies for a few Q lines were also reported by Ellis et al.\ \cite{ellis2012}.

The intensities of the lines are expressed in units of photons/cm$^2$/s,
normalised to the intensity of the brightest line, which is set to
10$^3$. The superscripts to the line intensities are used to flag
the reliability of the flux measurement. Their meanings are follows
\begin{description}
\item[$a$] : well detected line in a region free of telluric absorption.
The line is free from blending or can be de-blended.
The error on its relative flux is expected to be within 10\%\ r.m.s.
\item[$b$] : well detected line, but affected by some telluric absorption 
and/or blending and/or other problems. 
The error on its relative flux could be much larger than 10\%.
\item[$c$] : line flux poorly defined because the line is detected at
low s/n ratio or it is severely affected by telluric absorption or it
is strongly blended.
\end{description}

\subsection{Excitation and physical conditions of OH}

\begin{figure}
   \centering
   \includegraphics[width=\hsize]{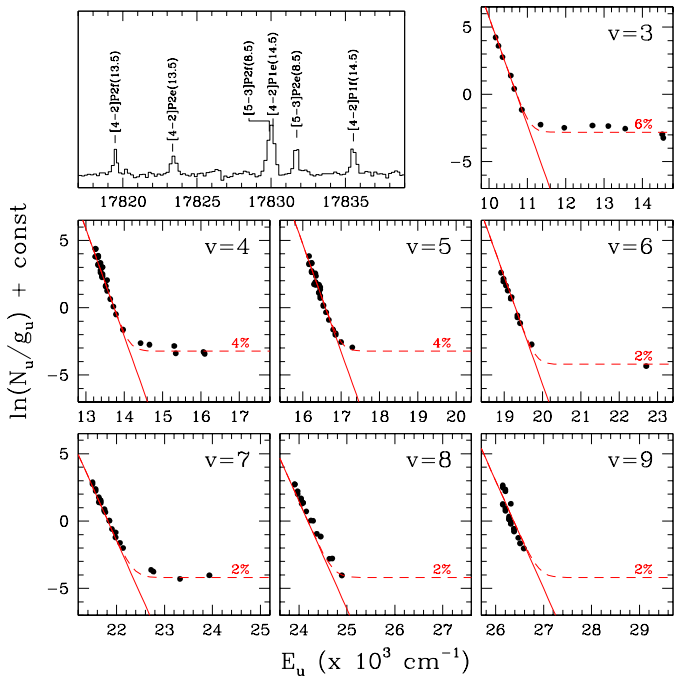}
      \caption{
   Relative populations of the OH levels. $E_{u}$ is the energy of 
   the upper level (units of $10^{3}$ cm$^{-1}$, equivalent to
   $1.99 10^{-13}$ erg) and $g_{u}$ is its
    statical weight. 
   The solid lines are
   the values expected for a thermal population with a vibrational
   temperature $T_{vib}$=9000~K and a rotational temperature $T_{rot}$=180~K.
   The dashed lines include a given fraction of ``hot'' molecules
   with $T_{rot}$=$T_{vib}$.
         The
         upper-left panel shows a region of the spectrum with some of the
         high excitation lines from ``non-thermal'' levels.
              }
         \label{fig_OH_levels_pop}
   \end{figure}

The physical conditions of the OH molecules can be determined by computing
the relative populations of the upper levels of the transitions, and comparing
them with thermal distributions. For this computation we used the
most up-to-date values of transition probabilities, i.e. those of
van der Loo \& Groenenboom \cite{vanderloo}. The results are shown
in Fig.~\ref{fig_OH_levels_pop} along with theoretical curves
(solid lines) for a thermalized population
with a vibrational temperature $T_{vib}$=9000~K and a much 
lower rotational
temperature $T_{rot}$=180~K, i.e. for typical excitation conditions
(see e.g. R2000). The most striking result is the flattening towards higher
energies in the observed distributions.
This large deviation from a thermal distribution is disclosed by the
measurement of lines arising from levels with rotational
quantum number as high as J=15.5. Some
of these lines are visible in the top panel of Fig.~\ref{fig_OH_levels_pop}
and in the central panel of Fig.~\ref{fig_unidentified}.

In Fig.~\ref{fig_OH_levels_pop} we also plot, as dashed lines, the
level population expected adding a certain fraction of ``hot'' molecules
with $T_{rot}$=$T_{vib}$.
The results of this simple model fit remarkably well the 
observations, but requires that the fraction of ``hot molecules'' increases
going to lower vibrational states. 
The results can be explained as follows. 
In the upper part of the mesosphere, the OH molecule is primarily formed by 
the reaction
$$ {\rm O}_3 + {\rm H} \rightarrow {\rm OH}^*(v\le9) + {\rm O}_2 $$
The freshly formed OH$^*$ molecule is in a excited vibrational and 
rotational state. At the typical densities of the mesosphere, collisional 
de-excitations within a given vibrational state are much faster
than spontaneous transitions (see e.g. Sharma \cite{Sharma85}). This process 
thermalizes the rotational levels of most OH molecules to the gas 
temperature. The ``non-thermal'' lines that we detect come from the small
fraction of OH$^*$ molecules which spontaneously decay before
thermalizing. 
The fact that this fraction increases for lower vibrational states
may indicate that the efficiency of collisional de-excitations decreases
for lower $v$'s.

There are a few points at low energies in the $v=9$ subplot of Fig.~\ref{fig_OH_levels_pop} 
that appear to be offset from the solid red line. These are the 
$\Delta v$=4 transitions which
will be discussed in the next Section.

\subsection{ Comparison with computed OH transition probabilities}

\begin{figure}
   \centering
   \includegraphics[width=\hsize]{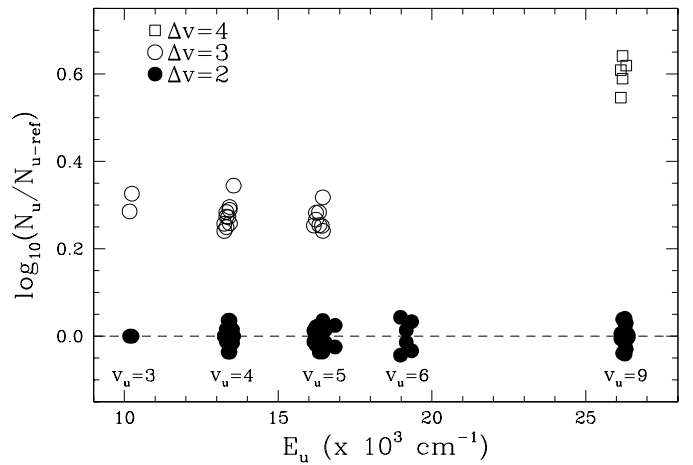}
      \caption{
 Comparison between observed and predicted flux ratios of
 OH lines from the same upper levels. Points with $N_u/N_{u-ref}$ far
 from unity (dashed line) indicate a discrepancy between observations
 and theoretical computations.
              }
         \label{fig_OH_lines_same_upper_level}
   \end{figure}

Another intriguing result follows from the analysis of the fluxes of
transitions arising from the same upper level. 
An excited molecule with a vibrational quantum number $v'$ and 
rotational quantum number $J'$ can spontaneously decay to a lower vibrational
state $v''$=$v'$-$\Delta v$ with rotational quantum numbers
$J''$=$J'$-1 (R line), $J''$=$J'$ (Q line)
and $J''$=$J'$+1 (P line). Therefore, depending on the value $J'$, there
are two (P+Q) or three (P+Q+R) lines for each $\Delta v$ band
(for a complete scheme of the OH transitions network see Fig.~2 of R2000). 
Since these lines are optically thin, their photon fluxes are simply 
determined by the population of the upper level $N_u$ times the 
transition probability $A_{ul}$, i.e.
$$ I = N_u A_{ul} $$
A convenient method to compare observations with theoretical
computations is to plot the value of $N_u$ derived from
different lines sharing the same upper level. The results are shown in 
Fig.~\ref{fig_OH_lines_same_upper_level}.
For each excited state, identified by its energy $E_u$, we include all
the lines with reliable flux measurements. The computed value of $N_u$ 
is normalised to the value derived from the brightest line
i.e.
$$ Y = \log(N_u/N_{u-ref})_i = \log(I/A_{ul})_i - \log(I/A_{ul})_{ref} $$
where the suffix $i$ refers to the line under consideration.
Values of $Y$ close to zero imply good agreement between observations
and theory. This is the case for all the points relative to the $\Delta v$=2
lines (filled dots in Fig.~\ref{fig_OH_lines_same_upper_level}) which
are distributed around the $Y$=0 line with a scatter compatible with
the observational errors.
 
The unexpected result is the systematic displacement 
of the other points. The $\Delta v$=3 lines are clustered around 
$Y$=0.3, while the $\Delta v$=4 transitions have an average
value of $Y$=0.6. In either cases the scatter of the points around their 
average values is compatible with the observational errors.
This result indicates that the computed
transition probabilities of the $\Delta v$=3 and $\Delta v$=4 lines are 
systematically underestimated by a factor of about 2 and 4, respectively.
Very similar results are found using the most recent transition probabilities,
published by van der Loo \& Groenenboom \cite{vanderloo}, and 
the older values, published by Mies~\cite{mies}.
The large discrepancy of lines with different $\Delta v$ is also evident 
in Table~\ref{tab_same_upper_level}, which lists the observed and 
predicted ratios for selected pairs of lines from the same upper level.

\begin{table}                                                                  
\caption{Photon flux ratios of OH lines from the same upper level}             
\label{tab_same_upper_level}                                                
\centering                                                                      
\begin{tabular}{|c|c|c|c|}                                         
\hline\hline                                                                    
 Line ratio\tablefootmark{1}  & Observed 
   & \multicolumn{2}{|c|}{Predicted\tablefootmark{2}} \\                   
\hline
[3-0]P1(2.5)/[3-1]Q1(1.5)  &  0.045   &   0.024 & 0.017 \\ \hline
[3-0]P1(3.5)/[3-1]Q1(2.5)  &  0.15    &   0.071 & 0.051 \\ \hline
[4-1]Q1(1.5)/[4-2]Q1(1.5)  &  0.11    &   0.063 & 0.060 \\ \hline
[4-1]P1(2.5)/[4-2]P1(1.5)  &  0.11    &   0.063 & 0.059 \\ \hline
[4-1]R1(1.5)/[4-2]R1(1.5)  &  0.12    &   0.063 & 0.061 \\ \hline
[5-2]P1(2.5)/[5-3]P1(2.5)  &  0.17    &   0.096 & 0.097 \\ \hline
[5-2]R1(1.5)/[5-3]R1(1.5)  &  0.17    &   0.098 & 0.10  \\ \hline
[5-2]Q2(0.5)/[5-3]Q2(0.5)  &  0.18    &   0.097 & 0.099  \\ \hline
[9-5]Q1(1.5)/[9-7]Q1(1.5)  &  0.21    &   0.058 & 0.075 \\ \hline
[9-5]P1(2.5)/[9-7]P1(2.5)  &  0.23    &   0.056 & 0.070 \\ \hline
    &  & & \\ \hline
[4-1]R1(2.5)/[4-1]P1(4.5)  &  0.55    &   0.54  & 0.61  \\ \hline
[4-2]P1(2.5)/[4-2]Q1(1.5)  &  0.76    &   0.76  & 0.74  \\ \hline
[5-2]Q2(0.5)/[5-2]P2(1.5)  &  0.50    &   0.49  & 0.50  \\ \hline
[5-3]R1(2.5)/[5-3]P1(4.5)  &  0.55    &   0.53  & 0.57  \\ \hline
[6-4]Q1(2.5)/[6-4]R1(1.5)  &  0.81    &   0.98  & 0.96  \\ \hline
[7-4]R1(2.5)/[7-4]P1(4.5)  &  0.55    &   0.56  & 0.61  \\ \hline
[8-5]P1(2.5)/[8-5]Q1(1.5)  &  0.73    &   0.74  & 0.72  \\ \hline
[9-5]R1(1.5)/[9-5]P1(3.5)  &  0.41    &   0.46  & 0.51  \\ \hline
[9-7]R2(1.5)/[9-7]P2(3.5)  &  0.49    &   0.50  & 0.51  \\ \hline
\hline                                                                          
\end{tabular}                                                                   
\tablefoot{\ \\
\tablefoottext{1}{ First group are pairs of lines with different $\Delta v$,
second group are pair of lines with the same $\Delta v$}\\
\tablefoottext{2}
{First entry is based on the transition probabilities of van de Loo 
\cite{vanderloo}, second entry is from the R2000 list, which is based on the
transition probabilities of Mies \cite{mies} 
}
}
\hspace{1.5cm}                                                                  
\end{table}

We were indeed very surprised of this result, to the point of questioning
the flux calibration of our data. In addition to double-checking all the
data reduction, we searched for other independent data which
could provide precise quantitative information on the relative fluxes of 
$\Delta v$=2,3,4 bands of OH. 
   \begin{figure}
   \centering
   \includegraphics[width=\hsize]{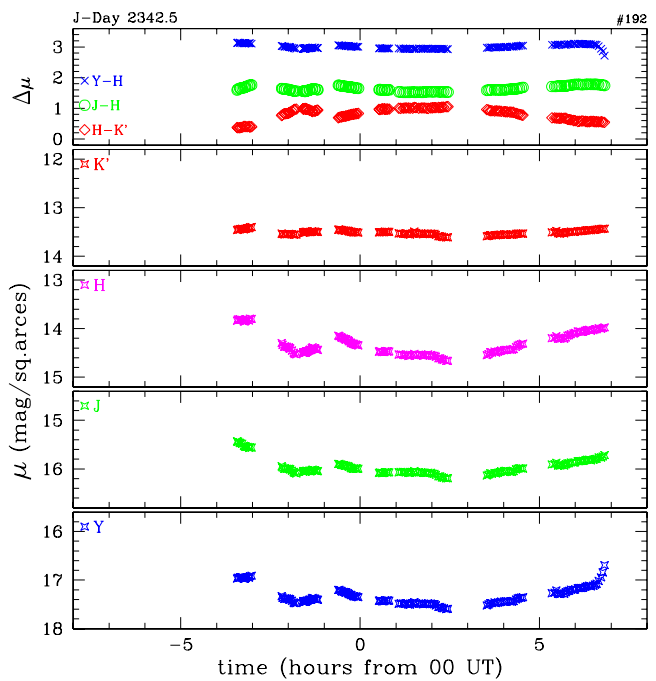}
      \caption{
  Variation of the airglow emission at La Palma during a dark night
 (moon 22\%, rising at 04:45 UT).
  The values of surface brightness and colors are expressed in
  Vega-magnitudes.
  The Y-H colors (uppermost points) are much bluer than predicted
  by theoretical OH spectra.
  See text for details
              }
         \label{Amici}
   \end{figure}

We used archive TNG data taken with the Amici disperser
of NICS, the TNG Near Infrared Camera Spectrometer
\cite{baffa2001}. These spectra simultaneously cover
the 0.9-2.5~$\mu$m wavelength range at a resolving power R$\simeq 50$.
Although the resolution is by far too low to measure the intensities of 
the single OH lines, the spectra can be conveniently used to derive the
integrated intensities and colours of the airglow within the infrared 
photometric bands. The results for a typical dark night are displayed in 
Fig.~\ref{Amici}. The emission in the Y (0.97--1.07~$\mu$m), 
J (1.17--1.33~$\mu$m), and H (1.48--1.78~$\mu$m) photometric bands is
dominated by airglow lines, while K' (1.95--2.30~$\mu$m) also includes
thermal emission from the telescope mirrors.
While the temporal variation in the airglow-dominated bands is quite large
(up to 1 magnitude), the colours are much more stable, and can be conveniently
used to compare with theoretical predictions. The J-H colour is difficult
to model, because of the strong contribution of O$_2$ lines in the J-band
(see Sect.~\ref{sect_O2}). The Y-H colour, instead, can be accurately 
modelled because OH accounts for most of the emission in both bands.
The airglow in the Y-band is mostly due to OH lines with $\Delta v$=3 
(3--0 and 4--1), while the H-band only contains OH lines
with $\Delta v$=2 (bands from 2--0 to 6--4). Therefore,
apart from minor effects related to small temporal variations of the 
vibrational and rotational temperatures, 
the Y-H colour should have a quasi-constant value which solely
depends on the relative transition 
probabilities between these vibrational bands. Using the published
transition probabilities we expect a photon flux ratio 
$N_\lambda$(H)/$N_\lambda$(Y)=10.8, equivalent to a colour Y-H=3.8.
The data in Fig.~\ref{Amici} confirm the predicted stability of the 
Y-H colour but, most important, shows that the airglow emission is 
0.8 magnitudes bluer than expected. In other words, the lines in Y band
are, on average, a factor of 2 brighter than predicted. This is the same
result that we find in the GIANO spectrum.

   \begin{figure*}
   \centering
   \includegraphics[angle=-90,width=0.85\hsize]{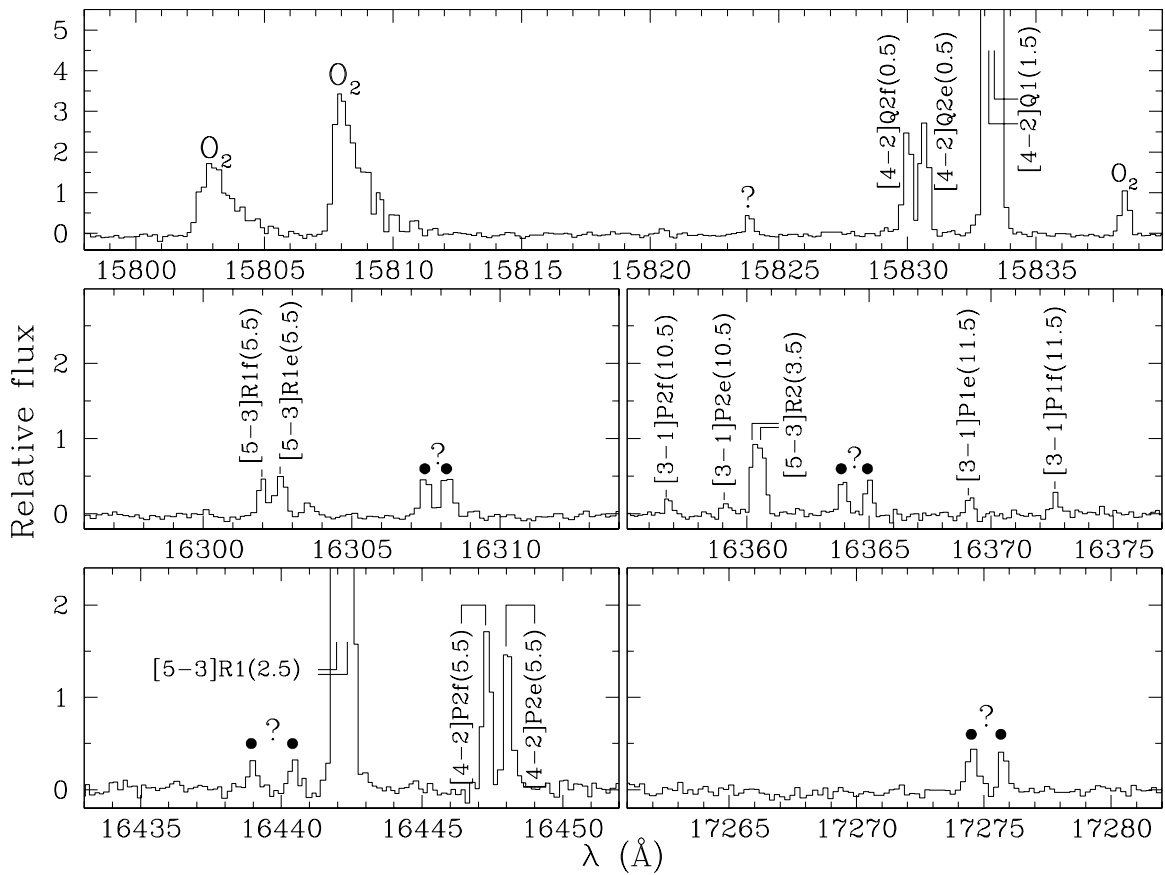}
      \caption{
Top panel: section of the GIANO spectrum including the band-heads of
O$_2$ (0,1) $a^1\Delta_g-X^3\Sigma^-_g$.
Lower panels: unidentified lines which appear as
$\Lambda$-split doublets. See text for details.
              }
         \label{fig_unidentified}
   \end{figure*}
\subsection{ O$_2$ and unidentified lines }
\label{sect_O2}
The GIANO spectrum includes about 150 lines which cannot be associated to
OH transitions. The measured line positions and fluxes are summarised
in Table~\ref{table_others_1}. The table lists the observed wavelengths 
(in vacuum) which are accurate
to 0.05~\AA\  r.m.s.  The relative photon fluxes 
and the accuracy flags ($a$=best, $c$=worst)  
are in the same unit as the OH lines (Sect.~\ref{OH_wavelengths_fluxes}).

All the brightest lines in the J-band 
are identified as roto-vibrational transitions of the O$_2$ 
(0,0) $a^1\Delta_g-X^3\Sigma^-_g$ band.
The observed wavelengths are equal, 
within the errors, to those listed 
in the HITRAN database (\cite{hitran2008}). 
Most of these lines 
are coincident with telluric absorption features, i.e. the O$_2$ lines
are optically thick.

The brightest emission features at longer wavelengths
are the broad features at 1.5803 and 1.5808~$\mu$m (see top panel of 
Fig~\ref{fig_unidentified}). 
These features are the band-heads of the
(0,1) $a^1\Delta_g-X^3\Sigma^-_g$ transitions of O$_2$. 
This is the first overtone of the (0,0) band discussed above.
Most of the lines in the
 1.56--1.61~$\mu$m range are also coincident with
O$_2$ transitions listed in the HITRAN database.
Interestingly, none of these emission lines is coincident
with telluric absorption features. Therefore, unlike the (0,0)~O$_2$
band, it seems that the (0,1)~O$_2$ lines are optically thin.

The remaining lines are  
unlikely to be associated to the O$_2$ band, 
because they are very far from the O$_2$ band-heads.
An intriguing
result is that several of these lines appear as closely
spaced doublets with equal intensities (see Fig.~\ref{fig_unidentified}).
In other words, they are very similar to all the $\Lambda$-split OH doublets
detected in our spectra. However, their wavelengths do not correspond
to any OH transition with $J_u\le19.5$ and $v_u\le10$.
The possibility that these doublets could be produced by OH isotopologues
(e.g. $^{18}$OH) should be investigated, but is beyond the aims of
this paper.

It is interesting to note that the integrated photon flux due to non-OH
lines is about 5\%\  of the total airglow line emission in the H-band.
Specifically, about 1.5\%\ is accounted for by the two band-heads 
around 1.58~$\mu$m, while the remaining 3.5\%\  is in isolated lines. 
This contribution is not necessarily
negligible and could complicate the design of airglow-subtraction
devices for astronomical instruments.

We do not include in the list the broad emission features at wavelengths 
beyond 2.3~$\mu$m, which are produced by absorption bands 
generated at relatively low heights in the atmosphere.

\section{Conclusions}

Using GIANO at the Telescopio Nazionale Galileo (TNG), we have obtained a
high resolution (R$\simeq$50,000) flux calibrated spectrum of 
the night airglow covering the 0.95--2.4~$\mu$m wavelength range. 
To the best of our knowledge, this is the first spectrum of this type
ever taken.

About 80\% of the detected lines can
be unambiguously identified as OH transitions. 
The observed wavelengths agree with those expected by the most accurate
molecular energy levels available in the literature 
(Abrams et al.\ 1994). 

The relative fluxes of OH are used to determine the physical conditions
of the emitting molecules. Most of the data are well fitted by a standard 
model, where the population of the vibrational states follows a Boltzmann
distribution at $T_{vib}$=9000~K, while the 
rotational levels within a given vibrational state are thermalized at 
$T_{rot}$=180~K.
However, we also 
detect lines from highly excited rotational levels. These reveal a
population of ``hot'' OH, with $T_{rot}\simeq T_{vib}$, which
accounts for a few \%\ of the total number of molecules.
This result indicates that the time-scales for the thermalization 
of the rotational levels are not quick enough to completely quench the 
emission from recently formed molecules in highly excited rotational states. 

Most surprisingly, the relative intensities of OH lines from the same upper
level show important discrepancies with what predicted by computed transition
probabilities. 

All the non-OH lines observed in the 1.2-1.3~$\mu$m range can be
identified as O$_2$ transitions within the 
(0,0) $a^1\Delta_g-X^3\Sigma^-_g$ band.
%
The remaining non-OH airglow lines are in the H-band (1.5-1.8~$\mu$m). 
Of these, about 2/3 are associated to the first overtone
of the same O$_2$ band, i.e. (0,1) $a^1\Delta_g-X^3\Sigma^-_g$ at 
1.58~$\mu$m.  
Interestingly, these lines are not coincident with telluric
absorption features, i.e. the lines are, most probably, optically thin.
The remaining lines, being far from the O$_2$ band-heads, are unlikely to
be associated to these band.


\begin{acknowledgements}
  Part of this work was supported by the grant TECNO-INAF-2011.
\end{acknowledgements}

\addtocounter{table}{-2}
\begin{longtable}{|r|r|r||r|r|r||r|r|rr|}
\caption{\label{table_oh_1} OH line wavelengths and relative photon fluxes}\\
\hline\hline                                                                    
 $\lambda_{vac}$ (\AA)\hfil &Iden\ \ \ \ \ \ \ \ & I\tablefootmark{1}\hfil &             
 $\lambda_{vac}$ (\AA)\hfil &Iden\ \ \ \ \ \ \ \ & I\tablefootmark{1}\hfil &             
 $\lambda_{vac}$ (\AA)\hfil &Iden\ \ \ \ \ \ \ \ & I\tablefootmark{1}\hfil \\ 
\hline
\endfirsthead
\caption{continued.}\\
\hline\hline                                                                    
 $\lambda_{vac}$ (\AA)\hfil &Iden\ \ \ \ \ \ \ \ & I\tablefootmark{1}\hfil &             
 $\lambda_{vac}$ (\AA)\hfil &Iden\ \ \ \ \ \ \ \ & I\tablefootmark{1}\hfil &             
 $\lambda_{vac}$ (\AA)\hfil &Iden\ \ \ \ \ \ \ \ & I\tablefootmark{1}\hfil \\
\hline
\endhead
\hline                                                                          
\endfoot
   9722.546 &[3-0]R1e(1.5)  &          &
   9740.794 &[3-0]R2e(0.5)  &          &
   9851.125 &[3-0]P2e(1.5)  &          &
\\ 
   9722.448 &[3-0]R1f(1.5)  &   18$^c$ &                                        
   9740.726 &[3-0]R2f(0.5)  &    9$^c$ &                                        
   9851.245 &[3-0]P2f(1.5)  &   16$^b$ &                                        
\\ \hline 
   9874.767 &[3-0]P1e(2.5)  &          &
   9897.437 &[3-0]P2e(2.5)  &          &
   9917.249 &[3-0]P1e(3.5)  &          &
\\ 
   9874.921 &[3-0]P1f(2.5)  &   38$^a$ &                                        
   9897.479 &[3-0]P2f(2.5)  &   11$^c$ &                                        
   9917.533 &[3-0]P1f(3.5)  &   41$^a$ &                                        
\\ \hline 
   9948.022 &[3-0]P2e(3.5)  &          &
   9959.328 &[9-5]R1e(1.5)  &          &
   9964.447 &[3-0]P1e(4.5)  &          &
\\ 
   9947.957 &[3-0]P2f(3.5)  &   11$^c$ &                                        
   9959.286 &[9-5]R1f(1.5)  &   21$^b$ &                                        
   9964.877 &[3-0]P1f(4.5)  &   23$^a$ &                                        
\\ \hline 
   9975.087 &[9-5]R2e(0.5)  &          &
  10002.832 &[3-0]P2e(4.5)  &          &
  10015.583 &[9-5]Q1e(1.5)  &          &
\\ 
   9975.079 &[9-5]R2f(0.5)  &    6$^c$ &                                        
  10002.642 &[3-0]P2f(4.5)  &   10$^c$ &                                        
  10015.527 &[9-5]Q1f(1.5)  &   44$^a$ &                                        
\\
\hline 
\hline                                                                          
\end{longtable}                                                                   
\tablefoot{\ \\
\tablefoottext{1}{I is the relative line
photons flux. The superscript
defines the accuracy ($a$=best, $c$=worst). See text for details.}
}

\newpage
\newcommand{\Otwo}{O$_2$}
\addtocounter{table}{1}
\begin{table*}                                                                  
\caption{Observed wavelengths and relative photon fluxes of 
  other airglow lines} 
\label{table_others_1}                                                          
\centering                                                                      
\begin{tabular}{|r|c|r||r|c|r||r|c|rr|}                                         
\hline\hline                                                                    
 $\lambda_{vac}$ (\AA)\hfil &Iden\tablefootmark{1} & I\tablefootmark{2}\hfil &                            
 $\lambda_{vac}$ (\AA)\hfil &Iden\tablefootmark{1} & I\tablefootmark{2}\hfil &                            
 $\lambda_{vac}$ (\AA)\hfil &Iden\tablefootmark{1} & I\tablefootmark{2}\hfil &                            
\\                                                                              
\hline                                                                          
12496.20 &    \Otwo      &   8$^a$ &                                               
12511.21 &    \Otwo      &  12$^a$ &                                               
12526.55 &    \Otwo      &  28$^a$ &                                               
\\ \hline 
12542.18 &    \Otwo      &  32$^b$ &                                               
12558.15 &    \Otwo      &  34$^b$ &                                               
12574.44 &    \Otwo      &  28$^b$ &                                               
\\ \hline 
12586.61 &  \Otwo    &   5$^b$ &                                               
12591.02 &    \Otwo      &  21$^b$ &                                               
12592.36 &    \Otwo      &   9$^a$ &                                               
\\ \hline 
12594.06 &    \Otwo      &   3$^b$ &                                           
12595.79 &    \Otwo      &   8$^a$ &                                            
12598.38 &    \Otwo      &  15$^b$ &                                            
\\ \hline 
12601.82 &    \Otwo      &  21$^b$ &  
12604.71 &    \Otwo      &  23$^b$ & 
12608.04 &    \Otwo      &  46$^b$ &
\\
\hline 
\hline                                                                          
\end{tabular}                                                                   
\tablefoot{ \ \\
\tablefoottext{1}{Identification based on
the HITRAN database.}\ \\
\tablefoottext{2}{I is the relative line
photons flux. The superscript
defines the accuracy ($a$=best, $c$=worst). See text for details.}
}
\hspace{1.5cm}                                                                  
\end{table*}


\begin{thebibliography}{}
  \bibitem[1994]{Abrams1994}
    Abrams, M.\ C., Davis, S.\ P., Rao, M.\ L.\ P., Engleman, R.\ Jr.,
   \& Brault, J.\ W.\ 1994, \apjs, 93, 351

  \bibitem[Baffa et al.\ 2001]{baffa2001} 
  Baffa, C., Comoretto, G., Gennari, S., et al.\ 2001, \aap, 378, 722

  \bibitem[2006]{Cosby2006}
    Cosby, P.\ C., Sharpee B.\ D., Slanger, T.\ G., Huestis, D.\ L.,
    \& Hanuschik, R.\ W.\ 2006, \jgr, 111, 2307

  \bibitem[2012]{ellis2012} Ellis, S.\ C., Bland-Hawthorn, J., Lawrence, J.,
   et al.\ 2012, \mnras, 425, 1682

  \bibitem[2003]{Hanushik}
    Hanuschik, R.\ W.\ 2003, \aap, 407, 1157

  \bibitem[1988]{Kaye}
     Kaye, J.\ A.\ 1988, \jgr, 93, 285

  \bibitem[1995]{Makhlouf}
    Makhlouf, U.\ B., Picard, R.\ H., \& Winick, R.\ J. 1995,
     \jgr, 100, 11289

  \bibitem[1993]{Maihara}
   Maihara, T., Iwamuro, F., Yamashita, T., et al.\ 1993, \pasp, 105, 940

  \bibitem[1974]{mies}
   Mies, F.\ H.\ 1974, J. Mol. Spec., 53, 150

  \bibitem[1992]{Oliva92}
  Oliva, E., \& Origlia, L.\ 1992, \aap, 254, 466

  \bibitem[2012a]{oliva_spie2012_1}
  Oliva, E., Origlia, L., Maiolino, R., et al.\ 2012, SPIE, 8446, 3TO

  \bibitem[2012b]{oliva_spie2012_2}
  Oliva, E., Biliotti, V., Baffa, C.,  et al.\ 2012, SPIE, 8453, 2TO

  \bibitem[1998]{Osterbrock98}
    Osterbrock, D.\ E., Donald, E., Fulbright, J.\ P., Cosby, P.\ C.,
    \& Barlow, T.\ A.\
    1998, \pasp, 110, 1499

  \bibitem[2011]{redman2011}
    Redman, S.\ L., Lawler, J.\ E., Nave, G., Ramsey, L.\ W., \& Mahadevan, S.\
    2011, \apjs, 195, 24

  \bibitem[Rothman et al.\ 2009]{hitran2008} Rothman, L.\ S., Gordon, I.\ E., Barbe, A., et al.\
        2009, JQS\&RT, 110, 533

  \bibitem[2000]{Rousselot00}
    Rousselot, P., Lidman, C., Cuby, J.-G., Moreels , G., \& Monnet, G.\
    2000, \aap, 354, 1134\ \ \ \  (R2000)

  \bibitem[1985]{Sharma85}
    Sharma, R.\ D.\  1985, Handbook of Geophysics, chapter 13 (Air Force
        Geophysics Laboratory, USAF)

  \bibitem[2007]{vanderloo}
    van der Loo, M.\ P.\ J., \& Groenenboom, G.\ C.\ 2007, J. of Chemical Physics,
   126, 114314

%
\end{thebibliography}
\end{document}